\documentclass[english,11pt,aps,prd,a4paper,preprintnumbers,floatfix,nofootinbib,showpacs,superscriptaddress, notitlepage]{revtex4-1} 
\usepackage[mode=buildnew]{standalone}
\usepackage[usenames,dvipsnames]{color}  
\usepackage{graphicx}
\usepackage{setspace}
\usepackage{caption}
\usepackage{subcaption}
\usepackage{amsmath}
\usepackage{amssymb}
\usepackage[colorlinks=true,citecolor=darkred,urlcolor=darkred, pdfborder={0 0 0}]{hyperref}
\usepackage[normalem]{ulem}
\usepackage{xcolor}
\usepackage{placeins}
\usepackage{mathrsfs}
\usepackage{verbatim} 
\usepackage{cancel} 
\usepackage{float} 

\usepackage{scalerel}
\usepackage{tikz-feynman}
\tikzfeynmanset{compat=1.1.0}
\usepackage{feynmp}
\usepackage{tikzsymbols}
\usepackage{multirow}

\makeatletter
\def\p@subsection{}
\makeatother

\definecolor{darkred}{rgb}{0.6,0,0}

\definecolor{linkcolor}{rgb}{0,0,0.5}

\def\gsim{\raise0.3ex\hbox{$\;>$\kern-0.75em\raise-1.1ex\hbox{$\sim\;$}}}
\def\lsim{\raise0.3ex\hbox{$\;<$\kern-0.75em\raise-1.1ex\hbox{$\sim\;$}}}

\def\beqn#1{\begin{equation}\label{#1}}
\def\eeqn{\end{equation}}

\def\beqa#1{\begin{eqnarray}\label{#1}}
\def\eeqa{\end{eqnarray}}

\newcommand {\ignore}[1]{}

\def\Z4{$Z_4$}
\def\O5{$\mathcal{O}_5$ }

\def\321{$\mathrm{SU(3) \otimes SU(2) \otimes U(1)}$ }

\usepackage{orcidlink}

\begin{document}

\title{\color{Blue} Implications of the First JUNO Results for Dirac Neutrino Texture Zeros}

 \author{Priya\orcidlink{0000-0003-1183-4441}}\email{kashyappriya963@gmail.com}
 \affiliation{Department of Physics and Astronomical Science, Central University of Himachal Pradesh
Dharamshala, India 176215}
\author{Ranjeet Kumar\orcidlink{0000-0002-7144-7606}}\email{kumarranjeet.drk@gmail.com}
\affiliation{Institute for Convergence of Basic Studies, Seoul National University of Science and Technology, Seoul 01811, Republic of Korea}
\author{Labh Singh\orcidlink{0000-0001-7704-726X}}\email{sainilabh5@gmail.com}
 \affiliation{Department of Physics and Astronomical Science, Central University of Himachal Pradesh
Dharamshala, India 176215}
\author{Surender Verma\orcidlink{0000-0002-5671-5369}}\email{s\_7verma@hpcu.ac.in}
 \affiliation{Department of Physics and Astronomical Science, Central University of Himachal Pradesh
Dharamshala, India 176215}

\begin{abstract}
  \vspace{1cm} 
  \noindent
Motivated by the first oscillation results from JUNO, we study the phenomenological viability of texture zeros in the Dirac neutrino mass matrix. The improved precision on the solar mixing angle $\sin^2{\theta_{12}}$ and the solar mass-squared difference $\Delta m_{21}^2$ provide a stringent probe for scrutinizing predictive texture zero frameworks. We perform a systematic scan of the allowed parameter space for two-zero textures, identifying sharp correlations among oscillation observables arising from the reduced parameter space.
Our analysis reveals that current JUNO measurements impose stringent constraints on the viable texture structures. In particular, although textures $C$, $A_2$, and $A_1$ were previously viable, current JUNO data strongly disfavor $C$, leaving only textures $A_2$ and $A_1$ compatible with the data. These findings underscore the remarkable sensitivity of Dirac texture zero scenarios to the solar sector.

\end{abstract}
\maketitle
\section{Introduction} \label{sec:intro}
The precision of neutrino oscillation parameters has improved significantly with the first data release from the Jiangmen Underground Neutrino Observatory (JUNO), enabling precise measurements of the solar sector. In particular, JUNO reports the solar mixing angle and solar mass-squared difference at the $1\sigma$ level as \cite{JUNO:2025gmd}
\begin{equation}
\sin^2\theta_{12} = 0.3092 \pm 0.0087, 
\qquad 
\Delta m_{21}^2 = (7.50 \pm 0.12)\times 10^{-5}\ \mathrm{eV}^2 .
\end{equation}
This corresponds to a significant improvement in precision by approximately a factor of $1.6$ relative to the combined constraints from previous reactor, solar, and long-baseline experiments.

These results establish the solar sector as one of the most precisely measured components of the leptonic mixing matrix, opening a new window for testing the flavor structures of viable neutrino mass models \cite{King:2017guk,Feruglio:2019ybq,Xing:2020ijf,Almumin:2022rml,Chauhan:2023faf,Ding:2023htn,Ding:2024ozt}. 
Traditional flavor symmetry based neutrino mass models are often successful in describing the leptonic mixing pattern, however, establishing simultaneous correlation with the measured neutrino mass-squared differences \cite{Super-Kamiokande:1998kpq,SNO:2002tuh} remains challenging\footnote{In recent studies, a scoto-seesaw type framework, augmented by flavor symmetry, has been successfully utilized to simultaneously account for the observed flavor structure and the two neutrino mass-squared differences \cite{Kumar:2023moh,Kumar:2024zfb,Borah:2024gql,Kumar:2025cte,Kumar:2025zvv,Nasri:2026nbf}.}. 
In this context, texture zero frameworks provide a minimal and predictive description of neutrino masses by imposing vanishing entries in the mass matrix~\cite{Frampton:2002yf,Desai:2002sz,Xing:2002ta,Guo:2002ei,Merle:2006du,Kageyama:2002zw,Xing:2003ic,Dev:2006qe,Dev:2009pa,Dev:2010if,Dev:2010vy,Verma:2020gpl,Lashin:2011dn,Deepthi:2011sk,Gautam:2015kya,Cebola:2015dwa,Cebola:2015dwa,Singh:2022ijf,Singh:2022tvz,Priya:2025khf}. Such textures naturally arise from underlying flavor symmetries or dynamical mechanisms and lead to characteristic correlations among neutrino masses and mixing parameters. Texture zero frameworks have been extensively explored for both Majorana and Dirac neutrinos. Nevertheless, the fundamental nature of neutrinos remains elusive. The observation of neutrinoless double beta decay would provide a definitive resolution to this question~\cite{Schechter:1981bd}, although current experimental searches have so far yielded inconclusive results~\cite{KamLAND-Zen:2022tow}.

In light of this, motivated by the first JUNO results \cite{JUNO:2025gmd}, we revisit the texture zeros of the Dirac neutrino mass matrix. The first JUNO results have prompted numerous theoretical studies~\cite{Chao:2025sao,Li:2025hye,Huang:2025znh,Ge:2025cky,Xing:2025xte,Chen:2025afg,Petcov:2025aci,Goswami:2025wla,Ding:2025dqd,Nanda:2025fvw,Shang:2026qkh,Zhang:2025jnn,He:2025idv,Jiang:2025hvq,Ding:2025dzc,Dutta:2026dzh,Kumar:2026qee}, including scenarios of texture zeros for the Majorana case \cite{Borah:2025vtn,Chambyal:2026fnq}. Out of the different possible texture zeros of the Dirac neutrino mass matrix, only the two-zero textures and one-zero textures remain viable, considering the charged lepton sector to be diagonal. For the one-zero texture, there are six possibilities and all of them remain allowed from the neutrino oscillation data.
While out of fifteen possibilities of two-zero texture, only three of them are viable \cite{Liu:2012axa}. These are referred to as the $A_1$, $A_2$, and $C$ scenarios \footnote{We are using the conventions given in Ref. \cite{Liu:2012axa}.}. Notably, for the $A_1$ and $A_2$ case, we have an interesting correlation among mass-squared differences ratio and mixing angles, given by (see Sec.~\ref{sec:text} for details),
\begin{align}
 A_1:   R_{\nu}= \frac{\Delta m^2_{21}}{|\Delta m^2_{31}|} \approx \frac{4 \tan^2{\theta_{23}}\sin^2{\theta_{13}}}{\sin{2 \theta_{12}}\tan{2 \theta_{12}}}, \quad A_2:  R_{\nu}= \frac{\Delta m^2_{21}}{|\Delta m^2_{31}|} \approx \frac{4 \cot^2{\theta_{23}}\sin^2{\theta_{13}}}{\sin{2 \theta_{12}}\tan{2 \theta_{12}}}. 
\end{align}
This offers direct testability of these scenarios, taking the latest JUNO results into account \cite{JUNO:2025gmd}. Furthermore, the $A_1$ and $A_2$ cases are only compatible with the normal mass ordering (NO), while $C$ remains consistent with both normal and inverted orderings (IO). Thus, the long term JUNO program to determine the neutrino mass ordering will provide a direct probe of these scenarios.

In this work, we examine the texture zeros of the Dirac neutrino mass matrix in light of the recent JUNO results \cite{JUNO:2025gmd}. We find that one-zero textures retain enough freedom of parameters and are not much affected by the latest JUNO results. In what follows, we focus our study on checking the viability of two-zero textures for the Dirac neutrino mass matrix. The normal ordering case for texture $C$ is disfavored by cosmological constraints on the sum of neutrino masses \cite{Planck:2018vyg,DESI:2024mwx}. In addition, the inverted ordering case is also ruled out once we impose the $\sin^2{\theta_{12}}$ constraint from JUNO results. 
Furthermore, for the $A_2$ texture, current global-fit analyses \cite{deSalas:2020pgw} indicate a clear preference for the upper octant of the atmospheric mixing angle ($\theta_{23} > 45^\circ$). In addition, once the latest JUNO constraint on $\sin^2{\theta_{12}}$ is taken into account, the allowed parameter space is significantly reduced. Consequently, only a narrowly restricted region remains compatible with the combined constraints (cosmology and JUNO), thereby enhancing the predictive power and testability of the $A_2$ texture. In contrast, the $A_1$ texture of the Dirac neutrino mass matrix exhibits a markedly different behavior under the same set of constraints. Among the phenomenologically viable two-zero textures, $A_1$ remains comparatively robust, with a sizable portion of its parameter space still consistent with both current cosmological bounds on the sum of neutrino masses and the JUNO sensitivity. This demonstrates the relative stability of the $A_1$ configuration, which retains substantial viability even in the presence of increasingly stringent experimental constraints.
Thus, JUNO results provide a stringent probe of two-zero textures of Dirac neutrinos. The texture $C$ is strongly disfavored and the parameter space of the textures $A_2$ is significantly constrained while texture $A_1$ remains viable by current cosmological and JUNO data.

The remaining structure of the paper is as follows. In Sec.~\ref{sec:text}, we discuss the two-zero textures of the Dirac neutrino mass matrix and provide the relation arising from them. In Sec. \ref{sec:res}, we present the numerical results, including possible correlations involving the solar mixing angle and examine their consistency with the recent JUNO results. Finally, we provide concluding remarks in Sec.~\ref{sec:conc}.

\section{Texture Zeros for Dirac Neutrinos} \label{sec:text}
We begin by revisiting the two-zero textures of the Dirac neutrino mass matrix \cite{Hagedorn:2005kz,Liu:2012axa,Borgohain:2020csn,Lenis:2023wyr}. Assuming a diagonal charged lepton mass basis, the Hermitian Dirac neutrino mass matrix $M_\nu$ is brought to diagonal form by the PMNS mixing matrix,
\begin{eqnarray}
U^{\dagger}_{\text{PMNS}}M_{\nu}U_{\text{PMNS}}=\text{diag}(m_1,m_2,m_3) \equiv M^{\text{diag}}.
\label{1}
\end{eqnarray}
In its standard parametrization, the PMNS matrix is expressed in terms of three mixing angles $\theta_{12},\theta_{13},\theta_{23}$ and a Dirac CP-violating phase $\delta$, given by
\begin{equation}
U =\begin{pmatrix}
U_{11} & U_{12}  & U_{13} \\
U_{21} & U_{22}  & U_{23} \\
U_{31} & U_{32}  & U_{33}\\
\end{pmatrix}=
\begin{pmatrix}
c_{12}c_{13} & s_{12}c_{13}  & s_{13}e^{-i\delta} \\
-s_{12}c_{23}-c_{12}s_{23}s_{13}e^{i\delta} & c_{12}c_{23}-s_{12}s_{23}s_{13}e^{i\delta}  & s_{23}c_{13} \\
s_{12}s_{23}-c_{12}c_{23}s_{13}e^{i\delta} & -c_{12}s_{23}-s_{12}c_{23}s_{13}e^{i\delta}  & c_{23}c_{13} 
\end{pmatrix} ,
 \label{2}
\end{equation}\\
where $c_{ij}=\cos{\theta}_{ij}$ and $s_{ij}=\sin{\theta}_{ij}$.
Employing Eqn. \ref{1}, we construct the Hermitian neutrino mass matrix in terms of the three leptonic mixing angles together with the CP-violating phase as
\begin{eqnarray}
M_\nu=UM^{\text{diag}}U^{\dagger},
\label{3}
\end{eqnarray}
where $M^{diag}=m_1,m_2,m_3$ contains neutrino mass eigenvalues.
In the SM, Dirac neutrino mass matrices are general complex $3\times 3$ matrices. 
Using the polar decomposition, any such matrix can be expressed as the product of a Hermitian and a unitary matrix, with the unitary factor absorbed into a redefinition of the right handed fields. Consequently, the Dirac neutrino mass matrix can be taken to be Hermitian without loss of generality. A $3\times 3$ Hermitian matrix contains six real parameters and three phases; weak-basis transformations reduce these to six real parameters and a single physical phase. This minimal set is sufficient to describe the three neutrino masses, three mixing angles, and the Dirac CP-violating phase. Imposing texture zeros further constrains the model, as two vanishing entries lead to two relations among physical parameters. The predicted values of these physical parameters are then examined in conjunction with the current oscillation data. Using Eqns. \ref{2} and \ref{3}, the six independent elements of $M_\nu$ can be written as
\begin{align}
M_{11} &= c_{13}^2 \left(c_{12}^2 m_1 + s_{12}^2 m_2 \right) + s_{13}^2 m_3 , \\[6pt]
M_{12} &= c_{12} c_{13} c_{23} ( -m_1 + m_2 ) s_{12}
+ c_{13} e^{-i\delta} \left( -c_{12}^2 m_1 + m_3 - s_{12}^2 m_2 \right) s_{13} s_{23} , \\[6pt]
M_{13} &= c_{13} c_{23} e^{-i\delta} \left( -c_{12}^2 m_1 + m_3 - s_{12}^2 m_2 \right) s_{13}
+ c_{12} c_{13} ( m_1 - m_2 ) s_{12} s_{23} ,\\ 
M_{22} &= c_{23}^2 \left( c_{12}^2 m_2 + s_{12}^2 m_1 \right)
+ s_{23}^2 \left[ c_{13}^2 m_3 + \left( c_{12}^2 m_1 + s_{12}^2 m_2 \right) s_{13}^2 \right]
+ 2 c_{12} c_{23} ( m_1 - m_2 ) s_{12} s_{13} s_{23} \cos\delta , \\
M_{23} &= c_{12}^2 c_{23} ( -m_2 + m_1 s_{13}^2 ) s_{23}
+ c_{23} \left( c_{13}^2 m_3 + s_{12}^2 ( -m_1 + m_2 s_{13}^2 ) \right) s_{23} \nonumber \\
&\quad
+ c_{12} e^{-i\delta} ( m_1 - m_2 ) s_{12} s_{13}
\left( c_{23}^2 - e^{2 i\delta} s_{23}^2 \right) ,
\end{align}
\begin{align}
M_{33} &= c_{23}^2 \left[ c_{13}^2 m_3 + \left( c_{12}^2 m_1 + s_{12}^2 m_2 \right) s_{13}^2 \right]
+ s_{23}^2 \left( c_{12}^2 m_2 + s_{12}^2 m_1 \right)
+ 2 c_{12} c_{23} ( -m_1 + m_2 ) s_{12} s_{13} s_{23} \cos\delta .
\end{align}
There are fifteen possible two-zero textures of $M_\nu$, however, only three are allowed by the current neutrino oscillation data \cite{ParticleDataGroup:2024cfk}, which can be represented as
\begin{equation}\label{twozero}
\begin{aligned}
A_1 &: 
\begin{pmatrix}
0 & 0 & X \\
0 & X & X \\
X & X & X
\end{pmatrix}, \hspace{0.3cm}
A_2 : 
\begin{pmatrix}
0 & X & 0 \\
X & X & X \\
0 & X & X
\end{pmatrix}, \hspace{0.3cm}
C : 
\begin{pmatrix}
X & X & X \\
X & 0 & X \\
X & X & 0
\end{pmatrix},
\end{aligned}
\end{equation}
where “$X$” denotes the non-zero element of the Hermitian Dirac neutrino mass matrix. In the present work, we investigate the phenomenological consequences of these three allowed two-zero textures in the Dirac neutrino mass matrix within the paradigm of JUNO's recent data. 
 In general, two-zeros in $M_\nu$ lead to two constraining equations given by
\begin{eqnarray}
(M_{\nu})_{ij}=0,\hspace{0.5cm}
(M_{\nu})_{kl}=0; \quad i,j\neq k,l.
\label{5}
\end{eqnarray}
 Therefore, using Eqn. \ref{3}, the two constraining equations can be written as
\begin{equation}\label{massratio}
\xi \;\equiv\; \frac{m_1}{m_3}
= \eta \,
\frac{
U_{i3} U^{*}_{j3} \, U_{\alpha 2} U^{*}_{\beta 2}
-
U_{i2} U^{*}_{j2} \, U_{\alpha 3} U^{*}_{\beta 3}
}{
U_{i2} U^{*}_{j2} \, U_{\alpha 1} U^{*}_{\beta 1}
-
U_{i1} U^{*}_{j1} \, U_{\alpha 2} U^{*}_{\beta 2}
},\hspace{0.3cm}
\zeta \;\equiv\; \frac{m_2}{m_3}
= \chi \,
\frac{
U_{i1} U^{*}_{j1} \, U_{\alpha 3} U^{*}_{\beta 3}
-
U_{i3} U^{*}_{j3} \, U_{\alpha 1} U^{*}_{\beta 1}
}{
U_{i2} U^{*}_{j2} \, U_{\alpha 1} U^{*}_{\beta 1}
-
U_{i1} U^{*}_{j1} \, U_{\alpha 2} U^{*}_{\beta 2}
}.
\end{equation}
Here, $\eta$ and $\chi$ represent the relative signs of $m_1$ and $m_2$ with respect to $m_3$, while the Dirac CP phase $\delta$ appearing in Eqn. \ref{massratio} is constrained by the requirement that all neutrino masses be real, thereby determining whether CP symmetry is conserved. For neutrino mass matrices with both texture zeros on the diagonal, CP violation is allowed: in this case $i=j$ and $k=l$, so Eqn.~\ref{massratio} remains real independently of the value of $\delta$; the texture $C$ permits CP violation. In contrast, textures $A_1$ and $A_2$ necessarily lead to CP conservation, since the real values of the neutrino masses force $\delta$ to take the discrete values $0$ or $\pi$. Further, the two mass-squared differences $\Delta m_{21}^2=m_2^2-m_1^2$ and 
$|\Delta m_{32}^2|=|m_3^2-m_2^2|$ together with the mass ratios
\begin{equation}
\left|\frac{m_1}{m_3}\right|\equiv r_1,\qquad \left|\frac{m_2}{m_3}\right|\equiv r_2,
\end{equation}
yield two independent determinations of the lightest neutrino mass $m_1$, given by
\begin{equation}
m_1^{(a)}=r_1\sqrt{\frac{\Delta m_{21}^2}{r_2^2-r_1^2}},\qquad
m_1^{(b)}=r_1\sqrt{\frac{\Delta m_{21}^2+|\Delta m_{32}^2|}{1-r_1^2}},
\end{equation}
respectively. Requiring consistency of the formalism, $m_1^{(a)}=m_1^{(b)}$, leads to the mass-ratio constraint
\begin{equation}\label{rnu}
\frac{\Delta m_{21}^2}{|\Delta m_{32}^2|}
=
\frac{r_1^2}{r_2^2-r_1^2}\left(1-r_1^2\right)
\equiv R_\nu .
\end{equation}
Further, using Eqn. \ref{massratio}, the analytical expression for the the $R_\nu$ for the texture $A_1$ and $A_2$ is given by \cite{Liu:2012axa}
\begin{align}
 A_1:   R_{\nu}= \frac{\Delta m^2_{21}}{|\Delta m^2_{31}|} \approx \frac{4 \tan^2{\theta_{23}}\sin^2{\theta_{13}}}{\sin{2 \theta_{12}}\tan{2 \theta_{12}}}, \quad A_2:  R_{\nu}= \frac{\Delta m^2_{21}}{|\Delta m^2_{31}|} \approx \frac{4 \cot^2{\theta_{23}}\sin^2{\theta_{13}}}{\sin{2 \theta_{12}}\tan{2 \theta_{12}}}. 
 \label{eqn:rnu}
\end{align}
Using the experimentally measured mass-squared differences $\left(\Delta m_{21}^{2},\hspace{0.05cm}\Delta m_{32}^{2}\right)$ together with the mass ratios $\left(r_1=\left|\frac{m_1}{m_3}\right|, \  r_2=\left|\frac{m_2}{m_3}\right|\right)$, the remaining neutrino masses can be written as 
\begin{eqnarray}
&&m_{2}=\sqrt{m_{1}^{2}+\Delta m_{21}^{2}};\hspace{0.1cm}
m_{3}=\sqrt{m_{2}^{2}+\Delta m_{32}^{2}}\hspace{0.2cm}\text{for NO   $(m_{1}<m_{2}<m_{3})$},\\ \nonumber
\text{and}\\ 
&&m_{2}=\sqrt{m_{1}^{2}+\Delta m_{21}^{2}};\hspace{0.1cm}
m_{1}=\sqrt{m_{3}^{2}+\Delta m_{32}^{2}-\Delta m_{21}^{2}}\hspace{0.2cm}\text{for IO   $(m_{3}<m_{1}<m_{2})$}.
\end{eqnarray}
In the next section, we present a numerical analysis of the $A_1$, $A_2$, and $C$ two-zero textures under the umbrella of global-fit and JUNO data.

\section{Numerical Results} \label{sec:res}

In our numerical analysis, we perform a random scan over the neutrino oscillation parameters by generating the mass-squared differences and mixing angles within their 3$\sigma$ ranges, as obtained from the latest global-fit results~\cite{deSalas:2020pgw}. In addition, the phenomenologically viable parameter space of the model is obtained by requiring the mass-ratio parameter $R_\nu$ to lie within its $3\sigma$ experimental range, $0.026 < R_\nu < 0.033$. The ordering of the neutrino masses, whether normal or inverted, is determined by $r_1$. For normal (inverted) ordering $r_1<1$ ($r_1>1$). The allowed parameter space is then projected onto contour plots corresponding to the global-fit regions of the oscillation parameters. Out of the fifteen possibilities of two zero textures, only three, $A_1$, $A_2$, and $C$ are viable from the global-fit results \cite{deSalas:2020pgw}.
Among them, textures $A_1$ and $A_2$, characterized by an off-diagonal vanishing matrix element, are CP-conserving and viable exclusively for normal ordering. In contrast, texture $C$ is consistent with both normal and inverted orderings and allows for CP violation. 
Next, we discuss the numerical results for textures $C$, $A_2$, $A_1$, and check their viability with the cosmology, global-fit results\footnote{It is to be noted that, in the NO, the two lightest neutrino mass eigenstates are separated by the smaller mass-squared difference, leading to a lower bound on the sum of neutrino masses of $\sum m_i \gtrsim 0.059$ eV. In contrast, for the IO, the smallest mass splitting occurs between the two heavier eigen-states, implying a stronger lower bound, $\sum m_i \gtrsim 0.10$ eV \cite{Gonzalez-Garcia:2021dve,DESI:2024mwx}.}, and the latest JUNO results.

\subsection{Texture $C$}
\begin{figure}[!h]
    \centering
    \includegraphics[scale=0.40]{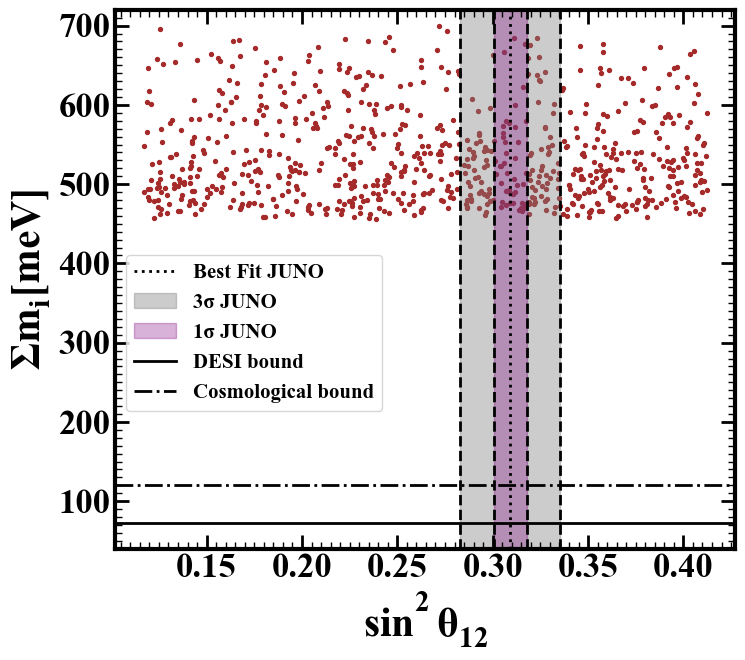}
    \includegraphics[scale=0.40]{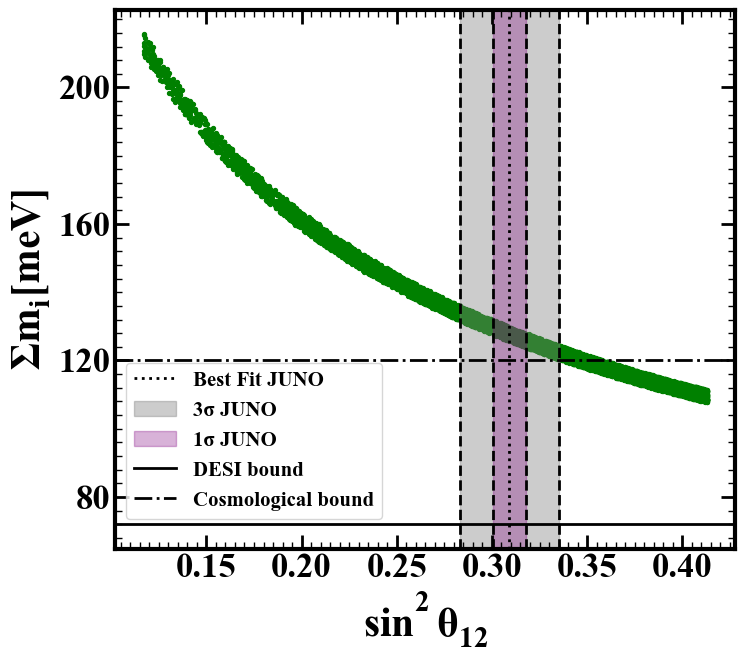}
    \caption{\textbf{Left (Right)} panel shows the correlation plots for the texture C scenario in normal (inverted) ordering between the sum of neutrino masses $\sum m_i$ and the mixing angle $\sin^2\theta_{12}$. The horizontal dot-dashed and solid black lines denote the upper bounds on $\sum m_i$ from Planck and DESI, respectively. The $1 \sigma$ and $3 \sigma$ regions of $\sin^2\theta_{12}$ from JUNO have been shown by purple and gray vertical bands, respectively.}
    \label{c1_ih}
\end{figure}

 We begin with texture $C$, which is allowed for both normal and inverted mass orderings. In the normal ordering, the $\sum m_i$-$\sin^2\theta_{12}$ correlation (left panel of Fig.~\ref{c1_ih}) shows that sum of neutrino mass $\sum m_i$ lies in the several-hundred-meV range, exceeding the current Planck ($<0.12$ eV) \cite{Planck:2018vyg} and DESI ($<0.072$ eV, 95\% CL) limits \cite{DESI:2024mwx}. 
Thus, the texture $C$ remains in strong tension with existing cosmological bounds. 
In this case, the $\sum m_i$ remains unaffected with JUNO constraints on $\sin^2\theta_{12}$, which is already severely constrained.

The right panel of Fig.~\ref{c1_ih} shows a strong dependence of $\sum m_i$ on $\sin^2\theta_{12}$ for texture $C$ in the inverted ordering, reflecting the restrictive nature of the underlying mass structure. The JUNO sensitivity constrains $\sin^2\theta_{12}$ to a narrow interval around its best-fit value, which in turn selects a region of parameter space with relatively large $\sum m_i$, thereby exceeding the cosmological bounds \cite{Planck:2018vyg}. In contrast, configurations that satisfy the Planck limit are realized only for higher values of $\sin^2\theta_{12}$, lying outside the JUNO preferred range. This separation between oscillation preferred and cosmologically allowed regions indicates a pronounced tension in the parameter space. The inclusion of the more stringent DESI bound further strengthens this conclusion, leading to a complete exclusion of the inverted ordering realization of texture $C$. Thus, texture $C$ is strongly disfavored by the combined constraints from cosmology and JUNO data.

\FloatBarrier

\subsection{Texture $A_2$}
\begin{figure}[!h]
    \centering
    \includegraphics[scale=0.40]{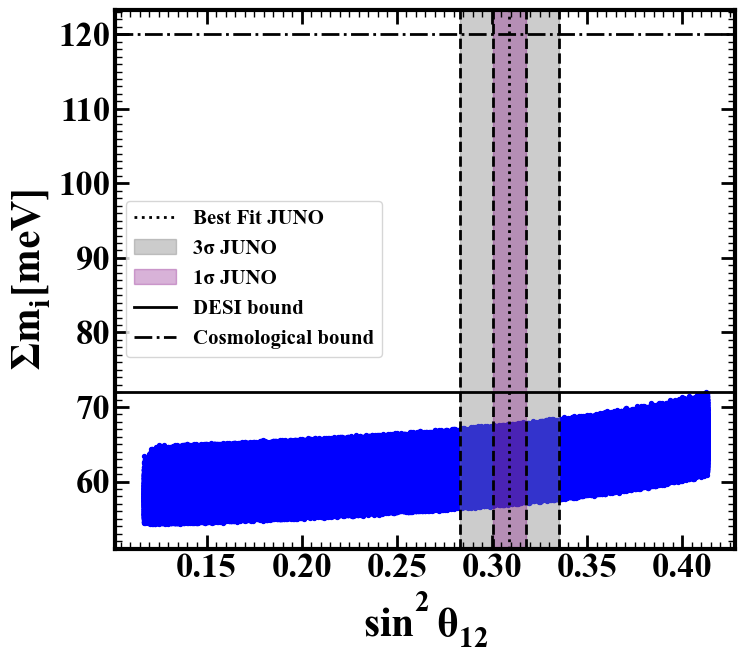}
    \includegraphics[scale=0.40]{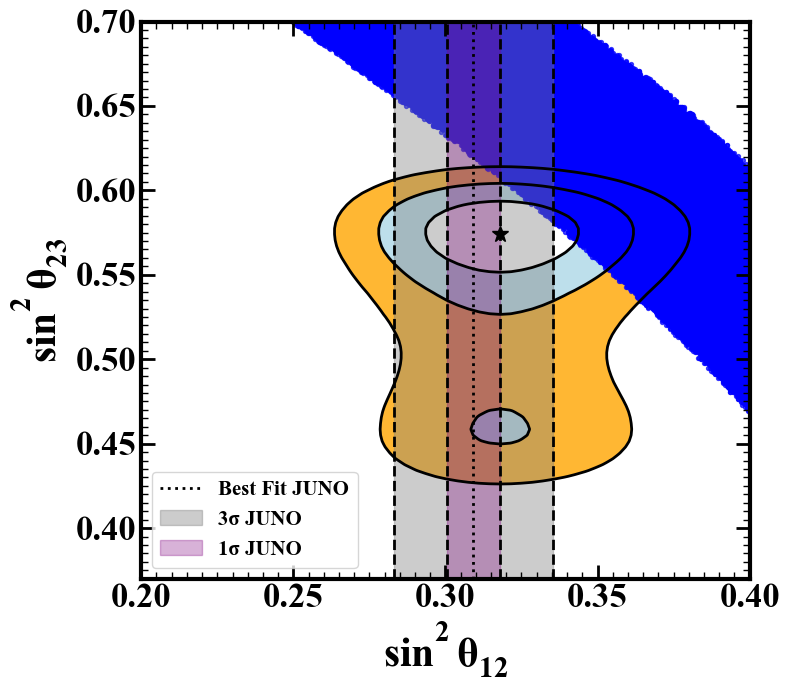}\\
         \includegraphics[scale=0.4]{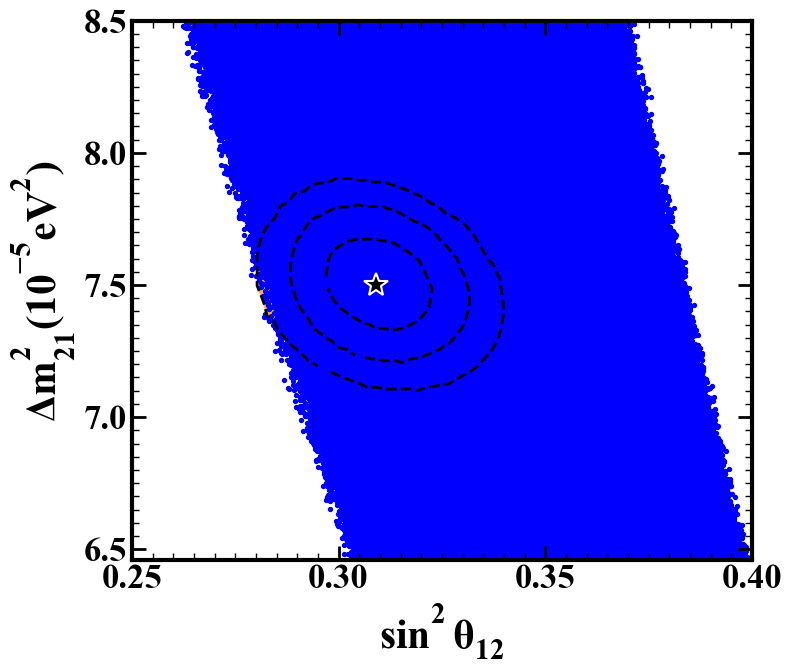}
    \caption{The correlation plots for texture $A_2$ illustrated in blue. \textbf{Upper left panel:} Sum of neutrino masses $\sum m_i$ vs solar mixing angle $\sin^2\theta_{12}$ has been shown. The horizontal dot-dashed and solid black lines denote the upper bounds on $\sum m_i$ from Planck and DESI, respectively. \textbf{Upper right panel:} This illustrates the correlation between the atmospheric mixing angle $\sin^2\theta_{23}$ and the solar mixing angle $\sin^2\theta_{12}$. The $1\sigma,2\sigma,3 \sigma$ contours in $\sin^2\theta_{23}$-$\sin^2\theta_{12}$ plane from global-fit have been shown for the comparison. In both panels, $1 \sigma$ and $3 \sigma$ regions of $\sin^2\theta_{12}$ from JUNO have been shown by purple and gray vertical bands, respectively. \textbf{Lower panel:} Correlation plots of solar parameters, $\Delta m^2_{21}$ and $\sin^2\theta_{12}$ for the texture $A_2$ corresponds to the JUNO contours. The $\boldsymbol{\star}$ represents the corresponding best-fit value.}
    \label{A2_1}
\end{figure}
%
We now discuss the numerical results for the texture $A_2$. This texture is only valid for normal ordering. Similar to texture $C$, we begin by checking its viability from cosmological constraints ~\cite{Planck:2018vyg,DESI:2024mwx}. In what follows, we present the correlation of the sum of neutrino masses $\sum m_i$ with $\sin^2\theta_{12}$. This has been illustrated with a blue band in the left panel of Fig.~\ref{A2_1}.
In this plot, we show constraints from cosmology (Planck and DESI) \cite{Planck:2018vyg,DESI:2024mwx} by horizontal lines, while the allowed region of $\sin^2\theta_{12}$ from the latest JUNO results \cite{JUNO:2025gmd} has been shown by a vertical band. We find that the texture $A_2$ remains consistent with the cosmology constraints. Moreover, the JUNO constraint substantially sharpens this prediction, compressing $\sum m_i$ into a narrow window, thereby enhancing the testability of the texture through future cosmological and precision oscillation measurements.
In the right panel of Fig.~\ref{A2_1}, we show the correlation between atmospheric mixing angle $\sin^2\theta_{23}$ and solar mixing angle $\sin^2\theta_{12}$. For the comparison, global-fit $1\sigma,2\sigma,3  \sigma$ contours and JUNO constraints by vertical band are shown. From the figure, it is evident that this texture intrinsically favors the upper-octant solution of the atmospheric mixing angle. Furthermore, the inclusion of the JUNO  constraints on $\sin^2\theta_{12}$ significantly restricts the viable region in parameter space. For example, while the upper $3\sigma$ limit of  $\sin^2\theta_{23}$ remains unaffected, the lower limit shifts upwards to 0.57 in the presence of the JUNO result. Next, we present the correlation between solar parameters, $\Delta m^2_{21}$ and $\sin^2\theta_{12}$ of neutrino oscillations shown in the bottom panel of Fig. \ref{A2_1}. The corresponding best-fit value is marked by the $\boldsymbol{\star}$. We show its comparison with the $1\sigma,2\sigma,3 \sigma$ contours of JUNO. We find that the texture $A_2$ remains consistent with the latest JUNO measurements. However, in order to illustrate the impact of improved precision in neutrino oscillation parameters, we consider a representative $1\sigma$ ranges of parameters for the texture $A_2$. The detailed analysis is presented in Appendix~\ref{appendix}. Our findings indicate that, while the texture remains consistent with global-fit constraints, it exhibits tension with the latest JUNO results.
%

\subsection{Texture $A_1$}
\begin{figure}[t]
    \centering
    \includegraphics[scale=0.40]{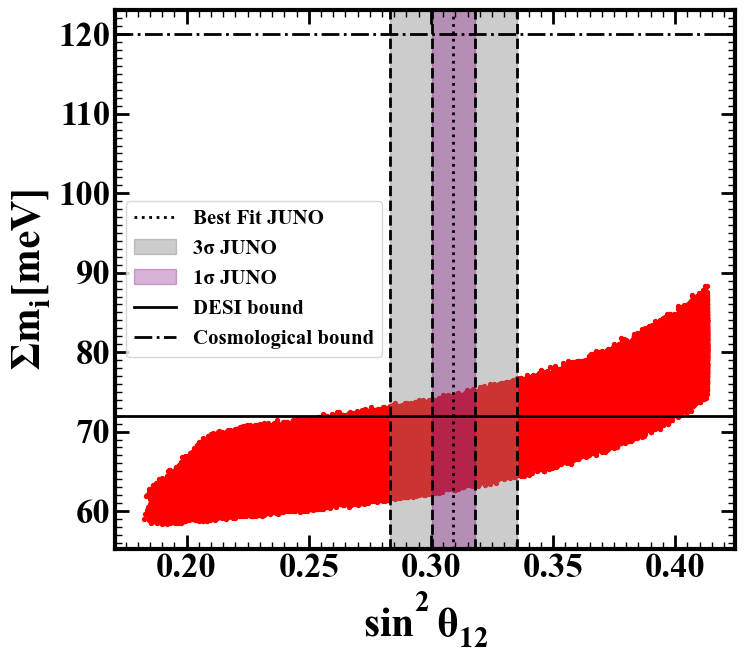}
    \includegraphics[scale=0.40]{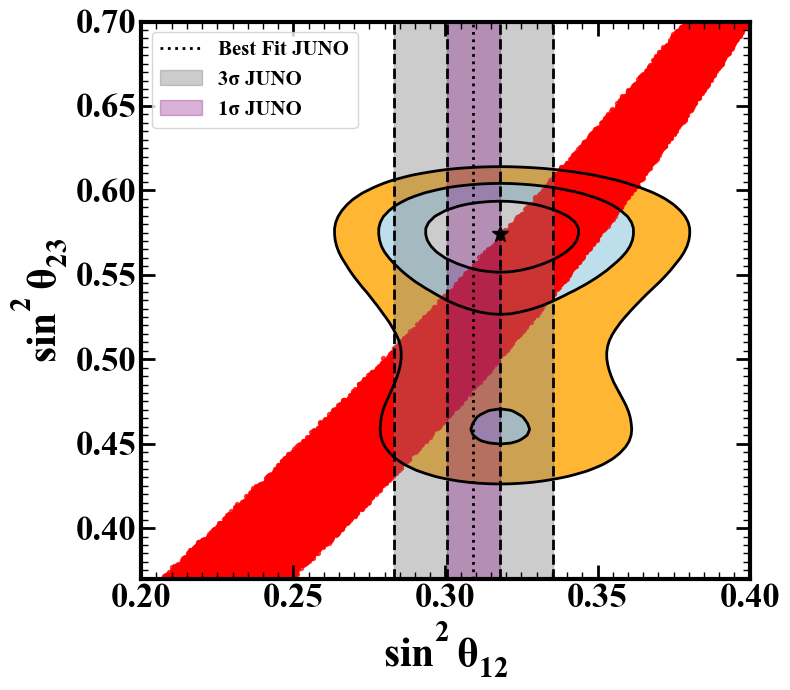}\\
    \includegraphics[scale=0.4]{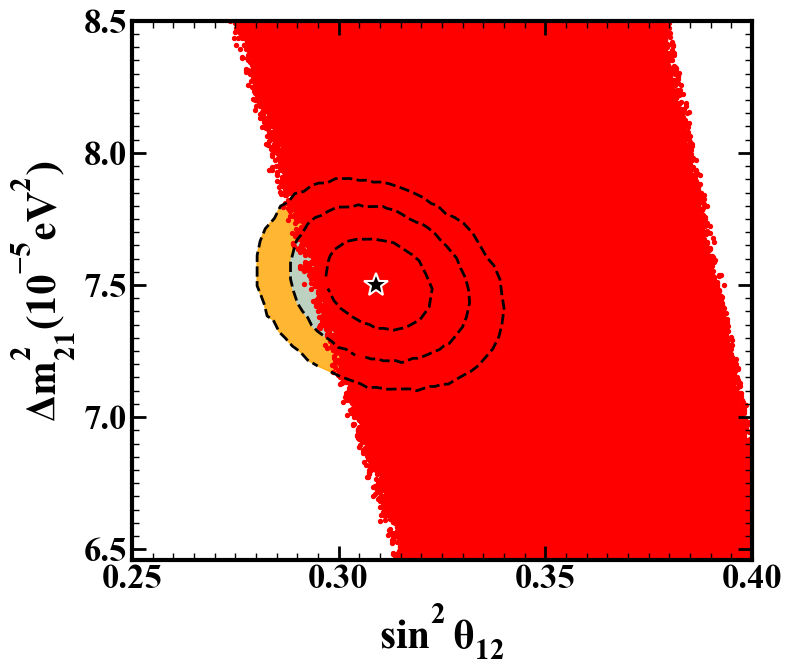}
    \caption{Correlation plots for the texture $A_1$ illustrated in red. \textbf{Upper left panel:} Sum of neutrino masses $\sum m_i$ vs solar mixing angle $\sin^2\theta_{12}$ has been shown. The horizontal dot-dashed and solid black lines denote the upper bounds on $\sum m_i$ from Planck and DESI, respectively. \textbf{Upper right panel:} This illustrates the correlation between the atmospheric mixing angle $\sin^2\theta_{23}$ and the solar mixing angle $\sin^2\theta_{12}$. The $1,2,3 \sigma$ contours of $\sin^2\theta_{23}$-$\sin^2\theta_{12}$ plane from global-fit have been shown for the comparison, where $\boldsymbol{\star}$ represents the best-fit value. In both panels, $1 \sigma$ and $3 \sigma$ regions of $\sin^2\theta_{12}$ from JUNO have been shown by purple and gray vertical bands, respectively. \textbf{Lower panel:} Correlation plots of solar parameters, $\Delta m^2_{21}$ and $\sin^2\theta_{12}$ for the texture $A_1$ corresponds to the JUNO contours. The $\boldsymbol{\star}$ represents the corresponding best-fit value.}
    \label{A1_1}
\end{figure}

We now delve into correlations of various observables of the neutrino sector for texture $A_1$. Similar to the textures $C$ and $A_2$, first, we discuss the viability of texture $A_1$ from cosmology. We present our results for the $\sum m_i$ vs $\sin^2\theta_{12}$ in the left panel of Fig.~\ref{A1_1}, shown in red (the remaining color code is the same as Fig.~\ref{A2_1}). 
Similar to the texture $A_2$, for $A_1$ we also find that, imposing the JUNO constraint of $\sin^2\theta_{12}$, the sum of neutrino masses narrows down. In addition, we show a correlation between the $\sin^2\theta_{12}$ and $\sin^2\theta_{23}$ in red (the remaining color code is the same as Fig.~\ref{A2_1}) in the right panel of Fig.~\ref{A1_1}. In contrast to the texture $A_2$, for $A_1$, both octants of $\theta_{23}$ are permitted. Next, we present the correlation between solar parameters, $\Delta m^2_{21}$ and $\sin^2\theta_{12}$ of neutrino oscillations. In the bottom panel of Fig.  \ref{A1_1}, we show its comparison with the $1\sigma,2\sigma,3 \sigma$ contours of JUNO. We find that the texture remains consistent with the JUNO data. This consistency persists even when the $1\sigma$ ranges of the neutrino observables in Eqn.~\ref{eqn:rnu} are considered, as illustrated in Fig.~\ref{A1_2} in Appendix~\ref{appendix}.

\section{Discussions and Summary} \label{sec:conc}
In this work, we revisit the texture zero structures of the Dirac neutrino mass matrix in light of the latest JUNO results. The one-zero textures, owing to their larger parameter freedom, are not much affected by the current JUNO data. In contrast, we find that two-zero textures are now tightly constrained. Although textures $C$, $A_2$, and $A_1$ were previously allowed, current JUNO data disfavor $C$, leaving $A_2$ and $A_1$ as the only viable textures consistent with JUNO measurements. 
Our results are obtained from Monte-Carlo sampling of the oscillation parameters within their $3\sigma$ allowed ranges. 
This enables us to examine the correlations among the mixing angles, the sum of neutrino masses, and solar oscillation parameters, as well as to assess the impact of JUNO precision on the phenomenological viability of the textures. The point-wise general remarks about the obtained phenomenology of allowed two-zero textures are as follows: 
\begin{itemize}
\item Texture $C$ remains formally allowed for both normal and inverted mass orderings. However, it faces severe constraints from cosmological observations. In the normal ordering, the predicted sum of neutrino masses is already well above the limits inferred from Planck and DESI, even in the absence of JUNO constraints. This places the scenario under strong tension with current cosmological data. In the inverted ordering, although a small region of parameter space can marginally satisfy the DESI bound, it is ruled out once the JUNO results are taken into account. Consequently, texture $C$ is strongly disfavored by the combined constraints from JUNO and cosmological observations.
\item Since the inverted ordering for texture $A_2$ is already excluded, we focus on its viability in the normal ordering. In this case, the sum of neutrino masses $\sum m_i$ remains consistent with cosmological data, and the atmospheric mixing angle $\theta_{23}$ is predicted to lie in the upper octant. Both observables exhibit strong correlation with the solar mixing angle, $\sin^2{\theta_{12}}$. Imposing the $\sin^2{\theta_{12}}$ constraint from JUNO data further restricts the allowed ranges of $\sum m_i$ and $\theta_{23}$.
We further examine this texture in the context of solar oscillation parameters, which have a strong correlation. The predicted parameter space in the $\Delta m^2_{21}$-$\sin^2{\theta_{12}}$ plane remains consistent with the latest JUNO measurements.
\item Similar to texture $A_2$, texture $A_1$ is also viable only for the normal ordering. We find that $\sum m_i$ and $\sin^2{\theta_{23}}$ possess strong correlations with $\sin^2{\theta_{12}}$ and remain consistent with current constraints. Furthermore, we analyze this scenario in light of the latest JUNO results in the $\Delta m^2_{21}$-$\sin^2{\theta_{12}}$ plane. The resulting parameter space exhibits a robust correlation. The behavior shows that this texture also remains compatible with both  JUNO measurements. 
\end{itemize}

 In summary, JUNO plays a crucial role in transforming two-zero Dirac neutrino textures from broadly allowed frameworks into highly predictive scenarios. While textures $A_1$ and $A_2$ remain viable, the texture $C$ is already under significant tension from cosmological constraints. These results emphasize the significant role of JUNO in probing Dirac neutrino texture zeros and offer a promising avenue for their potential exclusion in the near future.

\section*{Acknowledgments} 
\noindent
Priya would like to acknowledge the IUCAA for providing the HPC facility to carry out this work. R. K. would like to acknowledge the support from the National Research Foundation of Korea under grant NRF-2023R1A2C100609111. L. S. acknowledges the financial support provided by the Council of Scientific and Industrial Research (CSIR) vide letter No. 09/1196(18553)/2024-EMR-I. L. S. would also like to thank the organizers of ``Workshop on High Energy Physics Phenomenology 2025 (WHEPP-2025)" at IIT Hyderabad, where part of this work was discussed. The authors also acknowledge the Department of Physics and Astronomical Science, Central University of Himachal Pradesh, for providing the necessary facilities to carry out this work.

\appendix 
\section{} \label{appendix}

\begin{figure} [!h]
    \centering
          \includegraphics[scale=0.4]{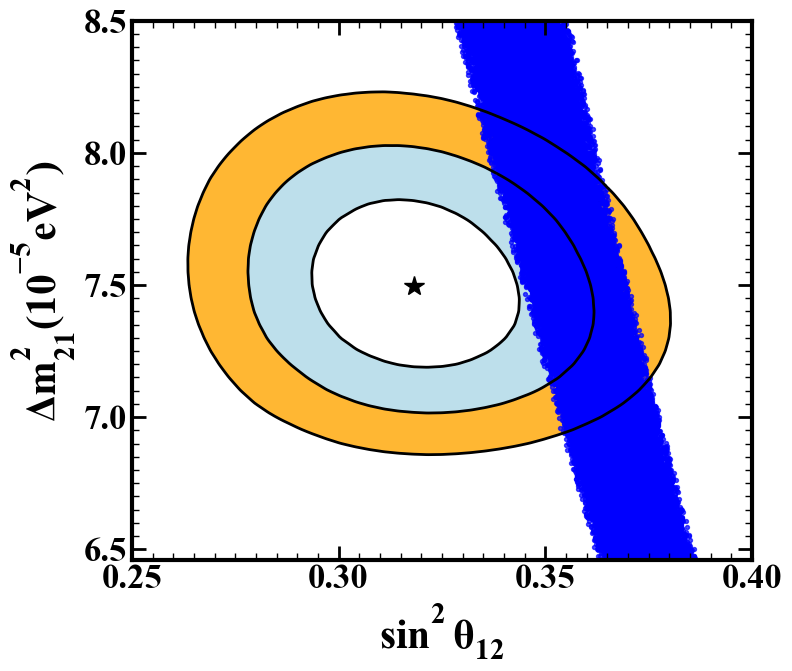}
        \includegraphics[scale=0.4]{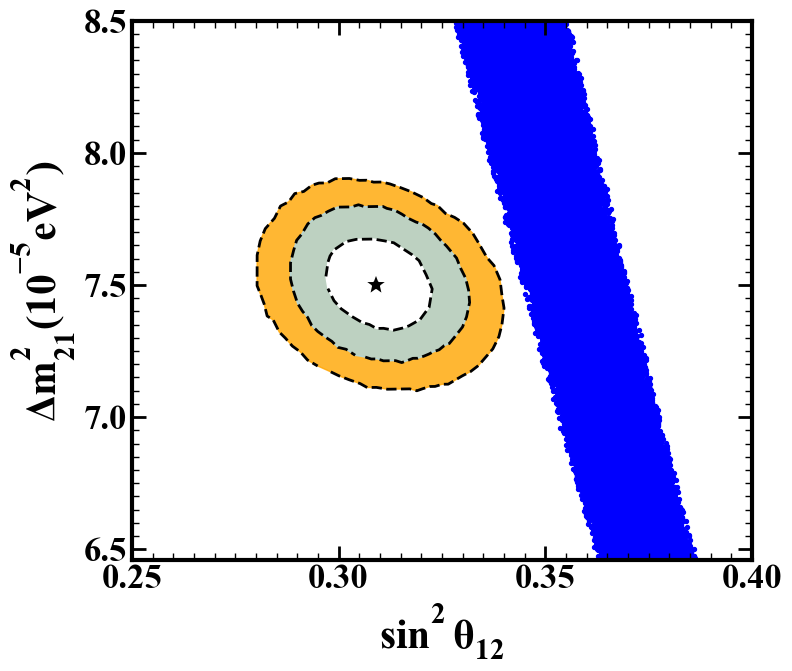}
\caption{Correlation plots of solar parameters, $\Delta m^2_{21}$ and $\sin^2\theta_{12}$ for the texture $A_2$ illustrated in blue. The \textbf{left} panel represents the comparison from the contours of global-fit data, while the \textbf{right} panel corresponds to the JUNO contours. The $\boldsymbol{\star}$ in both panels represents the corresponding best-fit value.}
\label{A2_2}
\end{figure}

Here, we present optimistic predictions for the textures $A_2$ and $A_1$ in view of prospective improvements in the precision of neutrino oscillation parameters. While both textures remain consistent with the current $3\sigma$ ranges, future measurements may provide tighter constraints, potentially enabling a more refined assessment of their viability.
We utilized the Eqn. \ref{eqn:rnu} to get the correlation between the  $\Delta m^2_{21}$ and $\sin^2\theta_{12}$. We have taken the $1\sigma$ values of all the parameters in Eqn. \ref{eqn:rnu} except $\Delta m^2_{21}$ and $\theta_{12}$. We present this scenario for texture $A_2$ in Fig. \ref{A2_2}. In the left and right panels of Fig.~\ref{A2_2}, we show its comparison with the $1,2,3 \ \sigma$ contours of global-fit and JUNO, respectively.
We find that this texture remains allowed from the global-fit (left panel) up to $2 \sigma$. While its prediction falls outside the $3 \sigma$ contours of the latest JUNO results (right panel). 
Thus, the latest JUNO results hint that the texture $A_2$ may be somewhat disfavored, with future precision measurements expected to further elucidate its viability.

\begin{figure}[!h]
    \centering
      \includegraphics[scale=0.40]{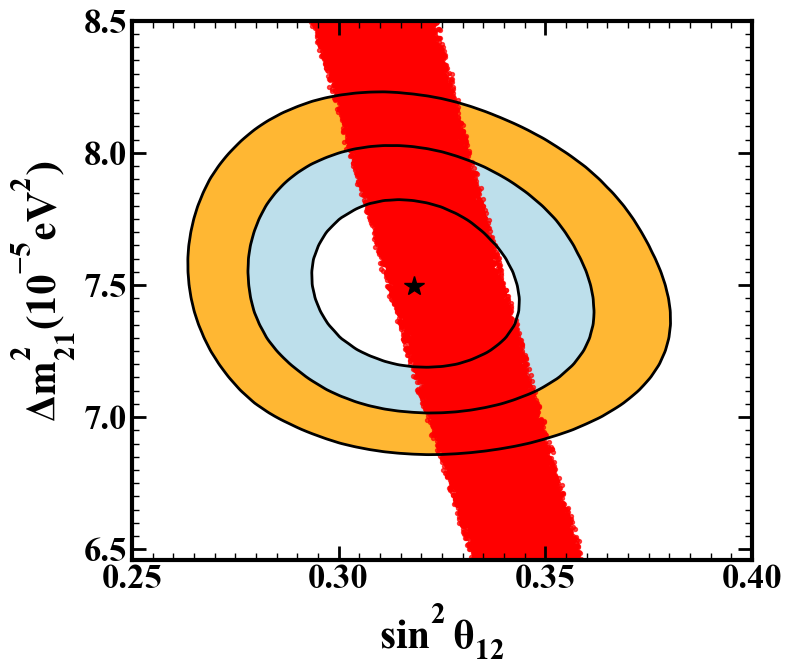}
        \includegraphics[scale=0.40]{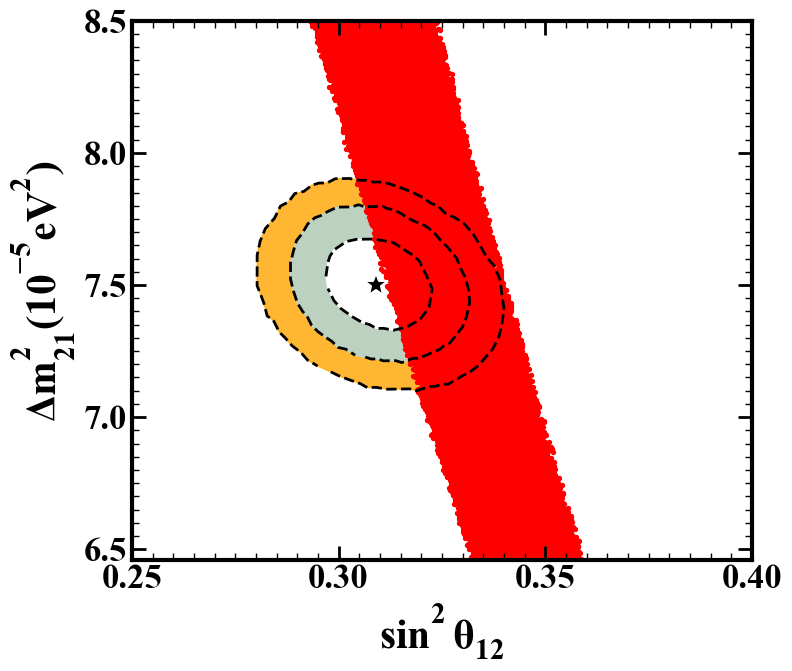}
\caption{Correlation plots of solar parameters, $\Delta m^2_{21}$ and $\sin^2\theta_{12}$ for the texture $A_1$ illustrated in red. The \textbf{left} panel represents the comparison from the contours of global-fit data, while the \textbf{right} panel corresponds to the JUNO contours. The $\boldsymbol{\star}$ in both panels represents the corresponding best-fit value.}
\label{A1_2}
\end{figure}
Similarly, we show this correlation for texture $A_1$ following Eqn. \ref{eqn:rnu}.
This has been presented in red in Fig.~\ref{A1_2}. In the left and right panels of Fig.~\ref{A1_2}, we show its comparison with the $1,2,3 \ \sigma$ contours of global-fit and JUNO, respectively. 
It is evident that the texture $A_1$ is viable from both the global-fit and JUNO results up to $1\sigma$. Therefore, texture $A_1$ remains well consistent with current measurements (discussed in the main text), and is expected to remain allowed in light of future high-precision data.

\FloatBarrier

\bibliographystyle{utphys}
\bibliography{references.bib} 
\end{document}